# THE FOUNDATIONS OF QUANTUM MECHANICS IN POST-WAR ITALY'S CULTURAL CONTEXT[1]

## Flavio Del Santo

### 1. The cultural roots of the renewed interest in the foundations of quantum mechanics

From a historiographical point of view, the change of fortune that the foundations of quantum mechanics (FQM) underwent over time is a fascinating case study of the interplay between science and the social-political environment within which it develops. Indeed, while FQM was highly debated in the early days of this theory and today are part of the mainstream research in physics, they have experienced a long period of near-oblivion. One of the reasons for the disappearance of FQM from physicists' agendas was the change of structures and priorities of scientific research during and after World War II, when physics became a tremendous driving force for new and wondrous applications, especially in the military sector. This period went down in history alongside the notorious expression "shut up and calculate!" (see Kaiser 2011).

In recent years, several works emphasized the role played by different cultural contexts in creating the conditions that led the field of FQM to experience a renaissance (Jammer 1974; Trischler & Kojevnikov 2011; Kaiser 2011; Freire 2014; Baracca et al. 2017; Baracca & Del Santo 2017; Freire 2019). This revival was a complex and discontinuous process prompted by a few pockets of resistance that arose in several different countries, often under the impulse of the social-political context. To illustrate the struggles of the physicists who pioneered the renaissance of FQM, Olival Freire Jr. has called them "the quantum dissidents" (Freire 2014). The latter encompass physicists like John Bell –who cultivated his interest on FQM as a mere "hobby", yet he put forward one of the most striking results of FQM, Bell's theorem (Bell 1964)– or David Bohm, who was led by his commitment to Marxism to formulate a non-standard realistic interpretation of quantum theory (Bohm, 1952).

Certain research environments reached a critical mass of physicists working on FQM already in the 1960s and 1970s, sometimes featuring very unconventional aims. It is the case, recently reconstructed by David Kaiser (2011), of the *Fundamental Fysiks Group* in Berkeley (California) which –in the late 1970s, in the context of a non-academic *New Age* mood– "planted the seeds that would eventually flower into today's field of quantum information science" (Kaiser 2011).

Another paradigmatic example occurred almost a decade earlier in Italy, where young, radically politicized physicists also revived the interest towards FQM. Being active at several Italian Universities (Bari, Bologna, Catania, Firenze, Rome and Trieste), they channelled the general political atmosphere of the 1968 left-wing struggles for dismantling the "capitalistic society", yet also within their field of expertise, namely physics. These radical physicists not only criticized the involvement of their colleagues in the military sector, but challenged the very contents of modern science, regarded as a product of the "capitalistic society". In such a way, the prevalent field of high energy physics –its methods and practices (*Big Science*)– became the target of harsh criticisms, too.

Quantum theory, instead, became the whipping post of these unsatisfied young physicists who saw in the newly introduced Bell Inequalities the tool to show its limits of validity, thus legitimising their overall critique of the scientific practice. These very ideological aims provided the motivation for the beginning of a systematic study of the state-of-the-art results in FQM that could have provided the opportunity for the introduction of a totally new physical framework. Although most of these attempts eventually turned out to be unsuccessful –the experiments on Bell's theorem confirmed the predictions of quantum mechanics– they opened a lasting period of intense research on FQM that involved dozens of Italian physicists and helped pave the way towards a recognition of FQM as a full-fledged field of physics. In fact, some of the Italian results acquired an international reputation, such as the celebrated "objective collapse model" developed by Giancarlo Ghirardi and his collaborators (Ghirardi et al. 1986) or the probabilistic solution to the measurement problem proposed by Marcello Cini (1983).

This chapter aims at reconstructing the cultural, political and ideological context that led to a revival of the foundations of quantum mechanics in Italy and at providing an overview of the main lines of research carried out from the post-war period until the end of the 1980s.

## 2. The years 1960s: The seeds of the Italian interest towards FQM

Like a great deal of intellectual activity in Europe, Italian physics was jeopardised by the scattering out of intellectuals from Europe after the advent of various forms of fascism and the consequent imposition of the racial laws (in Italy and in Nazi Germany). The outbreak of World War II, with the great involvement of physics in war technologies, completed the job of nearly annihilating any fundamental debate on FQM, not only in Europe but generally.

In the immediate post-war period, Edoardo Amaldi (1908-1989) contributed a great deal to rebuilding modern physics in Italy, with a strong focus on nuclear and high energy physics. Yet, it was especially Piero Caldirola (1914-1984) –professor of theoretical physics in Milan– that triggered new interest on FQM. Caldirola authored the entry "Quantistica, Meccanica" of the *Enciclopedia Italiana* (Caldirola 1961), that resulted in a quite comprehensive review of the main open conceptual issues in quantum physics.[2] Therein he acknowledged a proliferation of interpretational concerns, affirming: "the last decade [...] is characterised by an intense renewal of the attempts aimed at providing quantum mechanics with an interpretation different from that of the school of Copenhagen" (Caldirola 1961). Such an entry circulated as a booklet among the students of Caldirola's courses at the University of Milan and was to have a remarkable impact on some of those who were to play important roles in the following decade (such as Angelo Baracca and likely Ghirardi).[3]

Caldirola formed a school of theoretical physics, known as the "School of Milan", whose research was mainly focused on ergodic methods in statistical mechanics but, under his influence, some of his pupils brought together these statistical concepts with interest towards FQM. Indeed, in 1962, Adriana Daneri, Angelo Loinger and Giovanni Maria Prosperi developed a mathematical model that formalised –making use of the ergodic theorem from statistical physics– Bohr's explanation of the "collapse" of the quantum wave function due to an objective interaction between the macroscopic measurement apparatus and a microscopic (quantum) system (Daneri et al. 1962). This paper was timely and triggered a tremendous international debate, ultimately providing a concrete basis for the ongoing quarrel between Eugene Wigner and Léon Rosenfeld about the objectivity of the wave-function collapse.[4] Rosenfeld, who arguably was the staunchest amongst the supporters of Bohr's view, stated that "the Italian physicists have conclusively established the full consistency of the algorithm (of quantum mechanics), leaving no loop-hole for extravagant speculations" (Rosenfeld 1965).[5]

---

[2] In fact, a first work of Caldirola on FQM had appeared as early as 1957 in the philosophy journal "Il Pensiero" (Caldirola & Loinger 1957), but this likely had no resonance among the physicists.

[3] The fact that this book circulated among Caldirola's students in Milan is recalled by A. Baracca (see also Baracca et al. 2017).

As a further witness of the early Italian interest in quantum foundations, it is worth mentioning that, in 1967, Bruno Ferretti (1913-2010) –a former collaborator of Enrico Fermi, then professor of theoretical Physics in Bologna– organised a cycle of seminars on FQM.[6] This initiative involved the whole theoretical group in Bologna, where each physicist was requested to present a seminal paper on the recent developments of FQM. Among the young physicists involved were Baracca and Franco Selleri who were to play a pivotal role in the promotion of FQM and of other critical scientific activities.

Indeed, the following years saw a definitive break with the "old guard", and an unprecedented blossoming of the research on FQM in Italy. In the next section, we will show how this revival was, in Italy, deeply intertwined with the political radicalization of the young Italian physicists and their cry for an alternative science.

## 3. The 1968-effect: A rebirth of FQM in the context of left-wing political activism

### 3.1 The general social-political context

Before getting to the heart of the renewed interest that FQM underwent at the turn of the 1960s, we deem it necessary to quickly recall the general atmosphere of social and political struggles that characterized those years.

The student protests of Berkeley (1964), the "Prague Spring" (1968), the "French May" (1968), as well as the radicalisation of the labour struggles of the Italian "Hot Autumn" (1969) had a tremendous impact on society as a whole, and in particular on the themes that were set as a priority by the intellectuals. The active involvement of scientists in the dreadful Vietnam war, coming only a few decades after the awareness of the physicists' responsibility in the creation of the atomic bomb, sensitized many young scholars (see Moore 2013). This caused, throughout the 1960s, several outstanding incidents, such as the student strike at MIT (Boston) against military research, the intrusion of protesters into the meetings of the American Physical Society, and the boycotting of some talks given by eminent physicists, all members of the JASON advisory group.[7] This is the case of the Nobel laureate Murray Gell-Mann who was prevented from giving a talk at the Collège de France (Paris) in 1972 and again in CERN (Geneva) by young activist physicists. The same happened to Sydney Drell in Rome and to John Wheeler in Erice (Italy). Also the

---

[4] It would not be possible to enter here the details of this debate that has however been thoroughly discussed in (Freire 2014, chapters 4.4, 5.2). Nevertheless, it is worth mentioning that this involved some of the most preeminent physicists concerned with FQM in those years; among them – besides the aforementioned Rosenfeld and Wigner– Jauch, Shimony, Yanase, and Bohm.

[5] In the following years, Caldirola, too, helped a great deal to popularise in Italy the solution to the measurement problem proposed by his pupils. See, e.g., the paper (Caldirola 1965), and the textbook (Caldirola 1971), which devoted a whole chapter to the solutions of the measurement problem.

[6] The series of seminar is recalled by A. Baracca, V. Monzini, E. Verondini and S. Graffi who all participated to the initiative; communications to the author on June, 16th 2014.

[7] JASON is a group of distinguished physicists (today encompassing also other scientists), recruited to advise (originally by the American Institute for Defense Analysis, IDA) the US Government about science and technology, especially on classified military. It was established by the initiative of the distinguished physicists John Wheeler, Eugene Wigner and the mathematician Oskar Morgenstern in 1960 (See e.g., Finkbeiner 2006). JASON members (who, "[d]epending on how they're counted, they number between thirty and sixty at any given time", Finkbeiner 2006) played a significant role in the Vietnam war, when they promoted the "McNamara Line", an electronic barrier installed in South Vietnam to prevent infiltrations, and thus became the target of the new anti-war movements of activist scientists (See Vitale 1976).

Nobel Prize winner Wigner was openly criticised at the Varenna School in 1970 (see below), and in Trieste in 1972, where he even reacted displaying himself a banner with the words: "I am flattered by your accusations. They are compliments for me."[8]

Moreover, on the occasion of the first Moon landing in 1969, Marcello Cini (1923-2012) published a controversial essay, *Il Satellite della Luna* (Cini 1969), wherein he denounced the deep economic and military interests, disguised as scientific curiosity, behind the space race. This had a profound impact on the young critical physicists who became sensitive to the problems of scientists' responsibility (see Baracca et al. 2017).

Like in several other countries where left-wing political movements were gaining momentum, in Italy as well, students and young researchers did not limit themselves to refute physicists' involvement in military research, but initiated a critical analysis of the role of scientists in the (capitalistic) society, proposing concrete initiatives to change it. They sought in the methods and the contents of modern (capital-intensive) science some of the roots of the problematic issues of society. This atmosphere of unrest also seeped into the Italian Society of Physics (SIF) in October 1968, when exponents of the student movements interrupted the meeting of the Steering Committee (Baracca et al. 2017). Moreover, thanks to the initiative of Franco Selleri –who, as we will see in the next sections, was the real initiator of the renewal of the research on FQM in this ideological context– the Steering Committee of SIF voted in favour of renouncing to a substantial funding grant from the NATO for the year 1970.[9]

Following these premises, a new generation of radical, Marxist, Italian physicists started searching for alternative ideas and practices. It is precisely in this politicized context that the new wave of Italian research on FQM came about. It ought to be stressed that one aspect that renders the Italian case remarkably different from most of its international counterparts, seems to be that in Italy the political uneasiness that spread among physicists, enjoyed to a large extent the support of the institutions, in particular of SIF. Thus the usual clash between the academic establishment and the bottom-up movements of renewal was in Italy quite blurred. As we shall see, some of the protagonists of the Italian revival of FQM in those years played, on the one hand, the role of young, radical critics while, at the same time, being appointed to positions, such as members of the Steering Committee of SIF or the editorial board of its journal, *Il Nuovo Cimento*. It was not by chance that that journal became a pivotal international forum for the studies on FQM throughout the 1970s and 1980s.

### 3.2 Franco Selleri's early dissatisfaction with science and his pioneering initiatives

The undisputed protagonist of the revival of FQM in Italy –besides other political initiatives intertwined with physics– was Franco Selleri (1936-2013). After his graduation at the University of Bologna in 1958, Selleri made remarkable contributions to particle physics –which at that time was by far the most widespread field of physics– e.g., the introduction of the "peripheral one-pion model" (Bonsignori & Selleri 1960).

However, as early as 1965, Selleri found "evident that there were problems, fundamental problems in Physics".[10] But it was during a visiting professorship at the University of Gothenburg (Sweden), in 1966, that the major shift in Selleri's ideas took place, when he read the book *Conceptions de la Physique Contemporaine* by Bernard d'Espagnat (1965), about which he would later recall:[11]

---

[8] A full reconstruction of these episodes can be found in (Freire 2014, chapter 6) and (Vitale 1976).

[9] Eventually, despite the vote, the funding was accepted due to technical problems with the withdrawal of the application (Baracca et al. 2017).

[10] Franco Selleri interviewed by Olival Freire Jr. on June 24, 2003. Niels Bohr Library & Archives, American Institute of Physics (AIP), College Park, MD USA, www.aip.org/history-programs/niels-bohr-library/oral-histories/28003-1.

> It was a revelation. It was something fantastic to see how many problems were open in quantum mechanics. […] Reading that book was a great discovery. There was a new field possible to create and a lot of research to do. […] It was fascinating to see that so many possibilities were open. So it was clear that the Copenhagen approach was not unique, was not obligatory.[12]

Inspired By D'Espagnat's book, in 1969, Selleri published his first work on FQM: A short note entitled "On the wave function in quantum mechanics" (Selleri 1969a), whose preprint he had sent to Louis de Broglie, who, "manifest[ed] his great interest towards the research of the Italian physicist and the intent to intensify the contacts" (Nutricati 1998). That work is more the manifesto of a program than an actual research paper, but it is of great historical interest because it marks the moment in which realistic interpretations (in terms of hidden variables) entered the Italian physics landscape. Therein, Selleri maintained:

> Even though the theories of hidden variables are not completely developed, an important shift of philosophical attitude can be noticed: particles and waves are now objectively existing entities.

These positions were further developed by Selleri in the course of a series of lectures on *Quantum Theory and Hidden Variables* (Selleri 1969b) that he gave in Frascati (Italy) in June-July 1969. Within these lectures, Selleri formulated a "realistic postulate", which advocates a double ontological solution for particles and waves (i.e., they are both taken to objectively exist) and that would characterize his philosophical position ever after:

> An elementary particle is always associated to a wave objectively existing. […] 'Objectively' is taken to mean: independently on [*sic*] all observers. […] The wave has to be thought of as a real entity in some kind of postulated medium. Thus wave and particle are reminiscent of a boat in a lake. Boat and wave are both objectively existing and are found to be associated, in the sense that you cannot find a boat without a wave; the opposite is, however, possible.

Selleri was then able to put forward his ideas in a very institutional context, using –at least at a first stage– the reputation of SIF as a stepping stone to launch his unconventional views and reach a broad audience of Italian physicists. In fact, as a member of the Steering Committee of SIF between late 1968 and 1970, Selleri had the opportunity to promote critical initiatives aimed at rethinking the current scientific system. Despite the overall reluctance towards fundamental problems of physics at that time, Selleri proposed, in March 1969, to devote to FQM a one-week course of the celebrated "Enrico Fermi" Summer School in Varenna, organised annually by SIF, since 1953. In the minutes of one of their meetings, one can read that "after a wide discussion, the Steering Committee expresse[d] favourable opinion to the above-mentioned course and decide[d] to propose Prof. D'Espagnat as a director, highlighting of considering of interest a comparison of the opinions of Daneri, Prosperi, Loinger and Wigner, Bohm, de Broglie. As secretary of the course [was] proposed Selleri" (Baracca et al. 2017).

The acceptance, at that time, of such an official course on FQM by a national physical society appears to be remarkable and characteristic of the Italian case. In fact, as it was already mentioned, at that time there were just a few main spots of activities on FQM: In France the pupils of Louis de Broglie, in

---

particular Jean-Pierre Vigier (see below), were active to revive the idea of the "double-solution" (i.e. the corpuscular and undulatory natures of quantum particles). This had gathered new momentum after the proposed hidden variable model of David Bohm (1952), who, based in UK, was also one of the protagonists of the research on FQM. Vigier and Bohm were dissidents both in physics and in politics (being committed Marxists; see Freire 2011; 2014; 2019). However, despite their renown and the personal support of eminent physicists (de Broglie and Einstein, respectively) it does not seem that Vigier in France and Bohm in UK went any closer to breach in the institutions as Italian physicists did. Although it is not clear what are the factors that prepared the ground in Italy for this open-mindedness at the institutional level, this may again be related to the fact that quite radical political views were present and accepted in Italy, which, as a matter of fact, had the biggest national communist party of Western Europe. This political acceptance at the institutional level could thus have been reflected in the structure of scientific societies, in particular of SIF, at that time.

In fact, Selleri's success in this endeavour has surely been assisted by the presence in the board of SIF of yet another radical left-wing critical physicist, the above-mentioned Cini, who was to have a major impact on the following critique of science and later on FQM (see below). Moreover, the President of SIF at that time, Giuliano Toraldo di Francia (1916-2011), was an open-minded physicist, sympathetic towards left-wing ideas (though more institutional and less radical than Selleri and Cini), and very sensitive to ethical and philosophical problems of science.[13]

### 3.3 The Varenna School on Foundations of Quantum Mechanics of 1970[14]

The Varenna summer course on FQM opened on June 29th, 1970, and it constituted a meeting with a unique concomitance of features: the high profile of its teachers, the interest and the political unrest of its students, the open debates about interpretational topics –which had remained almost completely quiescent in public physics events since before World War II– and the first discussions about Bell's theorem. All together this makes clear why this School has been called "the Woodstock of the quantum dissidents" (Freire 2014).

The unconventional character of the meeting was already evident from the letter that the director of the School, D'Espagnat, sent to each participant beforehand (D'Espagnat 1971):

> Let me suggest to you the following agreement: that we should not take as goals the conversion of the heretic but rather a better understanding of his standpoint; that we should not suggest that we consider as a stupid fool anybody in the audience (lest the stupid fools should in the end appear clearly to be ourselves!); that we should try to cling to facts; and that nevertheless we should be prepared to hear without indignation very nonconformist views which have no immediate bearing on facts.

The school brought together 84 participants, of which the teachers were J. Andrade e Silva, J. Bell, B. D'Espagnat, B.S. De Witt, J. Ehlers, A. Frenkel, K.-E. Hellwig, F. Herbut, M. Jauch, J. Kalckar, L. Kasday, G. Ludwig, H. Neumann, G.M. Prosperi, C. Piron, F. Selleri, A. Shimony, H. Stein, M. Vigičić, E. Wigner, M. Yanase, and H.D. Zeh. (The proceedings contained also chapters from Louis de Broglie and David Bohm who however did not participate). The topics tackled in the school were organised into three main themes: (i) measurement and basic concepts, (ii) hidden variables and non-locality, (iii) interpretation and proposals. The only Italians among the teachers were Prosperi –who again proposed his solution to the measurement problem (see Sect. 2)– and Selleri, who presented a lecture to reaffirm his realistic position

---

and made use of Bell's inequalities (at that time virtually unknown), applying them to particle physics. In this work, Selleri also made evident his ideological motivations, by stating that

> practically nobody in the society of which the physicist is a part has any doubt about the actual existence of an objective reality outside of the observer. […] In this time where the social responsibility of the scientist is so strong, where the destruction or the survival of the world depends also on him, it is important to develop a science not in basic contradiction with the social reality. (in D'Espagnat 1971).

Overall, Varenna was a unique opportunity for those "dissident" physicists who, scattered around the world and working mostly independently, were timidly facing the main research directions of physics (at that time largely oriented towards practical applications) by exploring its neglected fundamental and philosophical aspects. At the genuine scientific level, "the school helped network these scientists, bringing together most of the physicists who would go on to contribute to the blossoming of this research in the 1970s." (Freire 2014). Worth mentioning it that Bryce de Witt announced in Varenna his conversion to Everett's *many-worlds interpretation* (as de Witt had named it and will have popularized ever after).

But there is more to that: The School was also a melting pot of ideas for the young physicists who, already sensitized by the general political climate, were eager to revolutionize their field of expertise both in its structures and scientific agendas. Among them were, besides Selleri: Angelo Baracca, Vincenzo Capasso, Gianni De Franceschi, Carlo De Marzo, Donato Fortunato, Gianni Mattioli, Alessandro Pascolini, Marta Restignoli, Luigi Solombrino, Tito Tonietti (a mathematician) and Livio Triolo. Many of them were to contribute to the research in quantum foundations in the following years. These attendants of the School, together with some of their international colleagues, gathered every evening after the lectures and engaged in critical discussions. Starting from contents of the courses, such as the paradoxes of quantum theory, they managed to relate these highly specialized and abstract concepts, such as the interpretation of the quantum formalism, with contemporary societal and political issues. A theme that became dominant was, in fact, the social responsibility of scientists in society. As a result, these politicized students produced a collective 12-page document, *Notes on the connection between science and society* (see Baracca et al. 2017), which was distributed to all the teachers and participants. The document aimed to raise awareness among the physicists about the oft-omitted relation between science and economical-political interests, and voiced the concept, which was to become influential in those years, of the *non–neutrality of science*:

> It is instead extremely important to realize that science is certainly not neutral [...]. The structures of a scientific theory reproduce the categories of the culture of the dominant classes. […] The limitation of the individual consciousness to laboratory activity, disregarding any judgement of the social application of research, results in indiscriminate support for all applications of science (e.g., atomic bomb, chemical and bacteriological warfare). […] The scientist's incapacity to control the product of his research facilitates its cultural manipulations and the creation of consensus. […] This mystification is formalized in the powerful theory ("scientism") which assumes the intrinsic ability of science to solve all the human problems. […] Now we conclude that a pre-decision is strongly needed concerning the social structure in which men live and act. That is, a pre-decision on the historical and social role of the scientist, on his responsibility, on the fact that no concept or activity is neutral."

In conclusion, the Varenna school of 1970 represented a gathering of physicists who would soon recognize themselves as a new community of "dissidents". Simultaneously, the *zeitgeist*, with an impetus of political renewal, entered the scientific debate and stimulated several young physicists to criticize the traditional values of scientific practice, as well as the content of research. Quantum foundations thus became the starting point for a concrete change.

### 3.4 Further critical activities about science in the "capitalistic society"

In the overall struggle for changing the scientific practice, the turn towards research in FQM was intertwined with a number of other critical activities.[15] Besides the mentioned critique to the practices in high energy physics, again triggered by Selleri, a major endeavor was carried out by the same young radical physicists who engaged in activities on the history of physics.

This came about once more within SIF: On the proposal of Selleri, SIF organized, in June 1972, a *Study Day on Science in the Capitalistic Society* in Florence. Whereas, the Varenna School of 1972 devoted a course to *The History of Twentieth Century Physics* (Weiner 1977), where prominent historians and physicists (such as Edoardo Amaldi, Paul Dirac, and Hendrik Casimir) held lectures before an audience which included several of the participants from Varenna 1970. Prompted by this, many of them engaged in professional research on the history of physics in the following years (such as A. Baracca, S. Bergia, G. Ciccotti, M. Cini, C. De Marzo and E. Donini). History was regarded as a means to understand the historical roots of modern physics and its positioning in the "capitalistic society", taking an evident inspiration from Marx's analysis of the political economy (*historical materialism*).[16]

This new approach soon entered a controversy with the established school of history and philosophy of science –also from a Marxist perspective– born around the figure of Ludovico Geymonat (1908-1991). However, this school –which had mostly a humanistic tradition and encompassed scholars such as Giulio Giorello, Enrico Bellone and Silvano Tagliagambe– voiced an analysis of the history of science based on Marx's *dialectical materialism* (see e.g. Bellone et al 1974). This approach was criticized by the physicists (turned historians), insofar as it seemed to imply absolute objectivity of science and therefore its legitimation as a neutral activity while neglecting the relationship between scientific development and societal needs.

## 4. The years 1970s-1980s: The blossoming of the activities on FQM in Italy

The many seeds of uneasiness towards the established scientific practice in the aftermath of the Varenna school on FQM led several Italian physicists to forever abandon the mainstream topics of research and pursue professional research in quantum foundations (as well as on the history of physics in many cases). Some of them created schools in their universities and established international collaborations.

The Italian journals of the group *Il Nuovo Cimento*, edited by SIF, played a pivotal role in disseminating work on FQM, also at the international level. Between 1969 and 1985, ca. 360 research papers on FQM authored by Italian groups appeared in these journals (Benzi 1988).[17] We provide here an overview of the main research programs on FQM in Italy in the decades of 1970s and 1980s.

### 4.1 Selleri's group in Bari and its collaborations

As already mentioned, Selleri had been the main catalyser of the political unrest and its intertwinement with physics that led to the Italian revival of FQM. In 1968, Selleri settled at the University of Bari, in the south of Italy, as the first professor ever of theoretical physics: A deliberate choice that allowed him full freedom for his research. In Bari, Selleri started teaching Quantum Theory in Winter 1968, supervised the first thesis on quantum physics and soon after some of his students fully embraced his ideas, making of Bari a centre for the research on FQM.[18]

---

[15] It lies beyond the scope of this chapter to thoroughly discuss the critique of science that involved the young Italian physicists in the 1970s. The interested reader is referred to (Freire 2014, Chapter 6.3), (Baracca et al. 2017, sections 9-12, 14) and (Baracca & Del Santo 2017).

[16] These activities led to many degree theses, research papers, dissemination articles, and books. See (Baracca et al. 2017) for a bibliography.

[17] It ought to be remarked that in 1969 –when the first Italian papers on FQM started to appear– the editorial board of *Il Nuovo Cimento* was composed only of members of the Steering Committee of SIF, including Selleri and Cini, whereas the editor in chief was SIF's president Toraldo di Francia. From the following year, while Toraldo di Francia remained the editor, the editorial board was vastly enlarged to include physicists of international renown –some of whom were also concerned with FQM– such as Hendrik Casimir, Raul Gatto, Josef-Maria Jauch, Giuseppe Occhialini, Bruno Pontecorvo, and Emilio Segré.

Selleri started prolific collaborations with some of his pupils, among whom were Vincenzo Capasso, Donato Fortunato, Augusto Garuccio, and Nicola Cufaro-Petroni. As acknowledged also by Kaiser (2011), "Selleri and his group in Bari, Italy, had been among the earliest and most active researchers on Bell's theorem anywhere in the world". However, it ought to be remarked that their aim was at first to disprove quantum mechanics by proposing experimental tests of Bell's inequalities (Selleri 1971; 1972). But afterwards –once the first experimental results corroborated quantum mechanics– Selleri and collaborators tried to show that the standard interpretation of Bell's inequalities had always been erroneous, in-so-far as it required additional assumptions.[19] A conviction that Selleri upheld until the end of his career: "Bell's inequality has never been checked experimentally. They have checked something else. That is to say another inequality based on local realism plus additional assumptions".[20]

Between 1979 and 1981, Selleri tried to stress the paradoxical aspects of quantum entanglement– again with the aim of proving a breakdown of quantum mechanics– by proposing protocols for superluminal signalling, which he presented in the course of scientific meetings, without however ever publishing his ideas.[21] These proposals were similar to those put forward by Nick Herbert from the aforementioned Fundamental Fysiks Group in the USA: the so-called "QUICK" experiment in 1979 and the "FLASH" experiment in 1981 (Herbert 1982; see also Kaiser 2011). Selleri became very sympathetic to these proposals, which were however conceived with the different spirit of actually achieving superluminal communication. After some correspondence, Selleri and Herbert met at a conference in Perugia and the former "was besotted with Herbert's latest proposal […]. More important, he pushed copies of Herbert's preprint on several other colleagues, and helped convince an experimental physicist from Pisa to mount a real test of Herbert's design" (Kaiser 2011).

All these proposed experiments assumed that it was possible to prepare a large number of identical copies of unknown polarized modes of a laser, which turned out to be a fundamental flaw. Yet, to show the infeasibility of these proposals, physicists had to come up with a solution that today goes under the name of "quantum no-cloning theorem": it is fundamentally impossible to perfectly copy an unknown quantum state (Wootters & Zurek 1982). Modern quantum cryptography relies directly on this theorem to guarantee communication security. However, as stressed by Kaiser (2011), the formulation of such a fundamental result was prompted by these idiosyncratic proposals.[22]

---

[18] The first thesis supervised by Selleri on Quantum mechanics was: Giglietto Antonio (1968), "Meccanica quantistica e processi markoviani", University of Bari, Italy. I am indebted to Luigi Romano for this reference (communication to the author on December 28th, 2019).

[19] Some works on this topic are (Capasso, Fortunato & Selleri 1973); (Fortunato & Selleri 1976); (Frotunato, Gariuccio & Selleri 1977) and (Selleri & Tarozzi 1978). We limit ourselves to this partial list, but the works of Selleri's group on Bell's inequalities continued until 1990. Comprehensive bibliographies of Selleri's work on FQM can be found in (Romano 2020) and (Nutricati 1998).

[20] Franco Selleri interviewed by Olival Freire Jr. in 2003; see footnote 10.

[21] Ghirardi recalled two preprints by Selleri on this topic: "Einstein locality and the quantum-mechanical long-distance effects," 1979, based on his presentation at a meeting in Udine, and a second preprint by Cufaro-Petroni, Garruccio, Selleri, and Vigier, 1980 (see Kaiser 2011). Herbert Pietschmann recalls that Selleri presented before him and Walther Thirring a proposal for superluminal communication during the leave of absence that Selleri spent in Vienna in 1980-1981 (Pietschmann interviewed by the author on 15-11-2016).

[22] The priority in the discovery of the no-cloning theorem is slightly contested. Usually attributed to Wootters and Zurek (1982) and independently to Diecks, Ghirardi, in his referee report of Herbert's FLASH experiment for the journal *Foundations of Physics*, in 1981, already provided a proof of the theorem that, however, remained unpublished. Moreover, it was recently shown that this theorem had already been proven in 1970 (Ortigoso 2018).

At the beginning of the 1980s, Selleri's group moved its focus toward experimental proposals for the detection of the hypothesized *empty waves* (Selleri 1982a; 1982b), (Andreade e Silva et al. 1983). In fact, as has already been mentioner, Selleri upheld a "dual" ontological solution for particles and waves (ideas very similar to those of de Broglie and the early Bohm). However, while in Selleri's view particles are always associated with a wave, there might also be waves that do not carry particles (empty waves). The proposed setup for their detection featured single photons sent through a beam-splitter. On the left branch a photodetector was locate, and on the right one a laser-amplifying tube. Selleri's idea was that after postselecting the cases when the photon gets detected on the left, the "ghost" photon on the right would still cause a (detectable) stimulated emission in the laser-amplifying tube. These proposals continued until 1990, but no experiments seem to have been carried out and the very feasibility of these proposals was questionable (Mückenheim 1988).

Moreover, Selleri spent a leave of absence at the University of Vienna in 1980-81. There, he lectured on FQM, and this plausibly had an impact on the Viennese students, and, as a matter of fact, on the young but influential physicist Roman U. Sexl who "followed [his] lectures in Vienna and invited [him] to write a book" on this topic.[23] The book (Selleri 1983) was indeed published (in German) in 1983 and became quite known in Austria and Germany.

Another important aspect of Selleri's impact on FQM was his ability in networking with international collaborators. Since 1969, Selleri established contacts with French Nobel laureate Louis de Broglie, with whom he had regular correspondence and several meetings; the whole Bari group had intense exchanges with the *Fondation de Broglie* in Paris, in particular with Georges Lochak, Pierre Claverie, Simon Diner and Olivier Costa de Beauregard.

Another lasting collaboration was established in 1978 with Vigier, who had a prime role in rebuilding the foundations of quantum mechanics since the 1950s.[24] This fully involved Cufaro-Petroni, who spent the years 1978-79 in Paris with Vigier, and Garuccio who worked in Paris with him in 1980. As a matter of fact, between 1979 and 1985, they published with Vigier 23 and 13 papers, respectively (see Baracca et al. 2017).

It is noteworthy that Vigier could count among his allies in the fight for reclaiming realism in quantum physics the influential philosopher Karl Popper, who had been opposing the Copenhagen interpretation since 1934 (See Del Santo 2019; 2020). Popper and Vigier together published a paper on quantum foundations that also involved Garuccio (Garuccio et al. 1981). In 1983, Vigier introduced Popper to Selleri (see Del Santo 2018), who promptly organized the conference *Open Questions in Quantum Physics*, to attract Popper to Bari. At the conference, Popper had the opportunity to present his variant of the EPR *gedankenexperiment*, usually referred to as just "Popper's experiment", and this marked his entry into the community of physicists concerned with FQM.[25] Allegedly, Popper's experiment was capable of empirically discriminating between a realistic interpretation of quantum physics and the standard Copenhagen interpretation, by violating Heisenberg's uncertainty principle. Also due to Selleri's effort, Popper's experiment became known internationally and triggered an intense, decade-long debate, also involving several Italian physicists. The experimentalist Francesco de Martini, in Rome, proposed to implement a variant that could be realized in his lab, but that did not persuade Popper. Selleri himself became convinced that the experiment was in principle infeasible due to the impossibility of preparing a suitable source of entangled particles (Bedford & Selleri 1985). Ghirardi rebutted Popper's proposal at a conference co-organised by Selleri in Cesena in 1985, calling it a "misunderstanding about the EPR analysis" (Ghirardi 1988).

---

[23] Franco Selleri interviewed by Olival Freire Jr. in 2003; see footnote 10

[24] Vigier was the assistant of de Broglie who convinced the latter to come back to his ideas of the pilot wave, after introducing to him the work of Bohm. Vigier has been one of the major influences in creating an international network between the quantum dissidents in the post-war era (see Besson 2018).

[25] The genesis of Popper's experiment and its reception, as well as the involvement of Selleri's group in its popularization, is discussed in (Del Santo 2018) and (Freire 2004).

Eventually, Popper's experiment was realized in 1999 at the University of Maryland, on Garuccio's suggestion, by Yanua Shih and Yoon-Ho Kim, who surprisingly found that Popper's predictions were right. They published this result in a controversial paper entitled "Experimental realization of Popper's experiment: Violation of the uncertainty principle?". It was only recently that a formal analysis of Popper's experiment showed that his proposal was in principle not able to test the Copenhagen interpretation (see Del Santo 2018).

Finally, since 1979, Selleri's group began a collaboration with the experimental group of Vittorio Rapisarda, from the University of Catania, for the design of an experimental test of Bell's inequality (Garuccio and Rapisarda 1981; see also Nutricati 1998). However, during the preparation of the experiment "Foca-2" (Falciglia et al. 1982), Rapisarda prematurely died in a car accident on a visit to Bari in 1982. Furthermore, the philosopher Gino Tarozzi co-organised with Selleri a series of international conferences on FQM that had a significant resonance: In Bari in 1983 (Tarozzi & van der Merwe 1985) and, in 1985, in Cesena (Tarozzi & van der Merwe 1988) and Urbino (van der Merwe et al. 1988).

Since the beginning of the 1990s, however, Selleri's interest towards FQM progressively faded, also due to the more and more evident empirical confirmations of quantum mechanical predictions in Bell-type experiments. However, some of his pupils kept working on quantum foundations and, in particular, Garuccio became specialized in quantum optics.

### 4.2 Giancarlo Ghirardi in Trieste

Giancarlo Ghirardi (1935-2018), who arrived a little later to the interests towards FQM, was not part of those radical physicists who questioned the validity of quantum mechanics but was directly influenced by them. In hindsight, his contributions have perhaps been the most outstanding among all the Italian studies on quantum foundations.[26]

As a student in Milan in the late 1950s, Ghirardi became aware of the fundamental issues of quantum physics quite early on, in a conference talk given by Prosperi. However, like many of his generation (including Bell), he pursued a safer career in mainstream physics until he got a permanent position in Trieste in the mid-1970s (see Kaiser 2011). It was again Selleri who stimulated him to enter a professional activity on FQM, as recalled by Ghirardi himself. However, Ghirardi could not fully endorse Selleri's radical ideological program: "I attended the lectures of Selleri and with a certain sympathy because I was left-wing. However, when I heard him say that Quantum Mechanics is a bourgeois science and ought to be rejected because it is unacceptable for a worker, then I felt very, very far away."[27]

After a few minor publications, such as on the implications of the Aharonov-Bohm effect (Ghirardi et al. 1976) and the stochastic interpretation of quantum mechanics (Ghirardi et al. 1978), it was in 1979 that Ghirardi started playing an important role in the international landscape. Indeed, Ghirardi and Weber (1979) found a flaw in Herbert's "QUICK" proposal for faster-than-light communication. Moreover, in his referee report of Herbert's FLASH experiment, dated April 22nd, 1981, Ghirardi, while recommending a rejection of the paper, proved the "quantum no-cloning theorem" (see footnote 22). Ghirardi also acknowledged the indirect influence of Selleri in developing the model that gave him international fame:

---

[26] A personal recollection of Ghirardi's activities on FQM can be found in (Ghirardi 2007).

[27] Email from Ghirardi to the author on 30-09-2016.

> [A]fter a seminar of Selleri it appeared clear to me that the core of the [measurement] problem was the superposition of the microstates that generates by reduction of the [wave] packet a statistical mixture. Since then I always had crystal clear in my mind that without breaking the linearity of the theory, we cannot get out of the contradiction.[28]

Indeed, Ghirardi and his collaborators Alberto Rimini and Tullio Weber developed a modification of the standard quantum formalism, adding non-linear terms to the standard Schrödinger equation that account for the collapse of the wave-function (Ghirardi et al. 1984; 1986), called the "GRW theory".[29] This was a first instance of an "objective collapse" model, in which the wave-function spontaneously undergoes a collapse at a random instant of time, according to an average time-rate proportional to the number of constituents of the system under study. As such, this provides a quantitative measure of the macroscopicity of a system and thus a neat solution to Schrödinger's cat paradox. The GRW theory was praised and popularized by Bell (1987), who connected Ghirardi's group with Philip Pearle. That collaboration led to the formulation of the more sophisticated model of "continuous spontaneous localization" (e.g., Ghirardi, et al. 1990).[30]

### 4.3 Angelo Baracca in Florence and Silvio Bergia in Bologna

Among the young physicists sensitized by the atmosphere of protest against the industrial complex practices in high energy physics and galvanized by the new trends in FQM, also thanks to the Varenna School of 1970, were Angelo Baracca and Silvio Bergia. They held professorships of theoretical physics at the Universities of Florence and Bologna, respectively. They soon started a collaboration which led to a first paper in 1974 (Baracca et al. 1974), wherein they attempted to operationally characterize the differences between entanglement and statistical mixtures. Following the Italian line of research initiated by Selleri, however, also their work was carried out with the expectation that an experimental test of Bell's inequalities could prove quantum mechanics wrong. In Summer 1974, they organized a national conference in Frascati, which brought together most of the Italian scholars that were entering this field, thus providing a first opportunity for this community to form an identity and exchange ideas.

Baracca, who became interested in Bohm's views at the Varenna School of 1970, visited Bohm at the Birkbeck College in London for two months in 1974. This collaboration led to a common paper (Baracca et al. 1975), wherein they put forward the idea that Bell's theorem "has no essential relationship to hidden variables, but rather that it is mainly significant as a test for whether or not the laws of quantum mechanics have to be extended in certain new ways". This again demonstrates the hope of employing Bell's inequalities to prove the limits of validity of quantum mechanics. Baracca also started supervising theses on FQM, of which the first was authored by Roberto Livi on Bell-type inequality for multivalued observables and proposed experimental tests using molecules (Livi 1977).

The collaboration between Baracca's and Bergia's groups continued until 1980 – extending the research on Bell's inequalities, and generalizing it to higher-dimensional variables (Baracca, Bergia & Restignoli 1974; Baracca et al. 1977; Baracca et al. 1978; Bergia et al. 1980). Later, Baracca moved his interest away from research in FQM, whereas Bergia and collaborators kept working on proposed tests of

---

[28] Email Ghirardi to the author on 30-09-2016.

[29] The paper (Ghirardi, Rimini & Weber 1986) gathered an impressive resonance, having by now been cited ca. 2700 times according to Google Scholar (last accessed July 11th, 2020).

[30] Continuous spontaneous localization had been independently formulated by Nicolas Gisin (1989). Moreover, the ideas of Ghirardi and co-workers have been further developed by some of their pupils, notably by Angelo Bassi, who recently put forward an experimental proposal to test collapse models against standard quantum mechanics (see https://www.nytimes.com/2020/06/25/magazine/angelo-bassi-quantum-mechanic.html?smid=em-share).

Bell's inequalities (Bergia & Cannata 1982; Bergia et al. 1985). They then proceeded to extend the ideas of Edward Nelson, according to which the evolution of a quantum mechanical system can be described in terms of stochastic processes (Bergia et al. 1988; 1989).

### 4.4 Marcello Cini in Rome

Marcello Cini, professor of theoretical physics in Rome, has been one of the central figures of the critiques to science in the 1970s (see Gagliasso at al. 2015; Aronova & Turchetti 2016). Cini was a dynamical intellectual, and, like Selleri, a radical leftist: A member of the Italian Communist Party since the 1940s, he became a dissident thereof when he co-founded the alternative communist newspaper *Il Manifesto*. In 1967, he also visited Vietnam during the war, as a member of the International Was Crime Tribunal (*Russel Tribunal*).

At the turn of the 1970s, Cini was member of the steering committee and vice-president of SIF, and became a reference for the Italian critique to science. He thus started working on the history of physics (with a Marxist approach) and on the social responsibility of scientists. His aforementioned work concerning the military interests in the space programs (Cini 1969) and his book against the neutrality of science, written in a Marxist spirit, *L'ape e l'architetto* (Ciccotti et al. 1976) had a tremendous impact on the critical scientists (see Aronova & Turchetti 2016).

Moreover, since the late 1970s, the focus of Cini's research in physics also moved towards FQM; an interest he kept until the end of his career. Similarly to the other Italians, Cini developed a critique against the Copenhagen interpretation of quantum theory, proposing a formal model that includes the measurement apparatus in the description of quantum mechanics (Cini et al. 1979; Cini 1983). Cini claimed that the fundamental problems of quantum mechanics stem from the idealization of isolated systems and the assumption that the measurement apparatus should lie outside of the domain of the theory (as a classical object). Indeed, he claimed that "the postulate of wave packet collapse, introduced as an extra assumption in quantum mechanics […], can be dropped and replaced by the Schrödinger time evolution of the total system object + apparatus". Although this approach was limited to particular cases only, it garnered a certain international interest, and even the prestigious journal *Nature* devoted a commentary in its "News and Views" section to it.[31]

### 4.5 Silvano Tagliagambe and quantum physics in the USSR

As a pupil of the aforementioned Geymonat school, Silvano Tagliagambe completed his studies in philosophy in 1968 with a thesis on Hans Reichenbach's interpretation of quantum physics. Between 1971 and 1974, he was in Moscow for a specialization on the philosophy of quantum mechanics under the supervision of the physicists Jacob Terletskij (1912-1993) and Vladimir Fock (1898-1974) and later with the philosopher Mikhail Omelayanovskij (1904-1979).

In the USSR, where Marxist ideology had pervaded all fields of knowledge, FQM had a two-sided tradition: The one voiced by Terletskij saw in the Copenhagen interpretation a despicable form of idealism, incompatible with materialism, while the opposite position –of which Fock was the most illustrious exponent– regarded Bohr's principle of complementarity as an expression of Marx' dialectical materialism (see Freire 2011). Thanks to these interactions, Tagliagambe soon became an expert on the philosophy of quantum mechanics in the Soviet context. In 1972, he edited the book *L'interpretazione materialistica della meccanica quantistica: Fisica e filosofia in URSS* (Tagliagambe 1972), which represented an exceptional testimony of the philosophical debate on FQM in the USSR. In his introduction, Tagliagambe stressed that "the volume […] has first and foremost an informational aim, being in its scope to give an idea to the Italian reader of the multitude of works that the Soviet scholars are carrying out […] on the most difficult problems of philosophy of science, and in particular of philosophy of physics" (Tagliagambe, 1972). Moreover, the volume also aimed at voicing the applicability of the methods of dialectical materialism (explicitly referring to Lenin's understanding thereof) to modern science. The volume, after a foreword by Geymonat and a historical preface by Tagliagambe, collected essays (translated into Italian by Tagliagambe himself) authored by seven physicists and sixteen philosophers on the occasion of various conferences held in the USSR between 1966 and 1971. Remarkably, the book also contains the first published version, in a language different from Russian, of the memoir that Fock wrote about the discussions he had with Bohr in Copenhagen in 1957. Therein, he supports his "compatibilist" view between the Copenhagen interpretation and dialectical materialism, stating: "Bohr's thought was always deeply dialectical […]. Such dialecticism was not 'spontaneous': Bohr told me that he studied dialectics in his youth and he always had held it in high esteem." (Fock, in Tagliagambe 1972).[32]

---

[31] "Uncertainties about uncertainty principle". *Nature* 302, 377 (1983).

However, this book came about within the already mentioned diatribe between historical and dialectical materialism in the context of the history of physics in Italy and presumably did not have an impact on the Italian physicists concerned with FQM.[33]

### 4.6 The Erice School of 1976

Among the academic events devoted to FQM organized in Italy in the late 1970s and 1980s, stands the international workshop "Thinkshop on Physics", held in Erice in April 1976. The School was directed by Bell and D'Espagnat, and also Selleri and his pupils participated in the event. If Varenna 1970 had represented a milestone in the legitimation of FQM, Erice provided the first opportunity for a new research community concerned with FQM to recognize itself. On this note, John Clauser, who was among the first to work on Bell's inequalities and a pioneer of modern foundations of quantum physics, later recalled that "the sociology of the conference was as interesting as was its physics. The quantum subculture finally had come out of the closet." (Clauser 1992).

It is also remarkable that Anton Zeilinger –who was to become one of the highest authorities in quantum foundations and quantum optics– became aware of the topics of FQM in Erice; in his words: "There, I heard for the first time about Bell's theorem, about the Einstein-Podolsky-Rosen paradox, about entanglement, and the like." (Bertlmann & Zeilinger 2013).

### 4.7 Other activities

We have until here focused on those lines of research which have been more evidently inspired by the politicized spirit of the time. As a matter of fact, FQM flourished in Italy throughout the 1970s and 1980s and it would be impossible to provide here a complete list of all the Italian research programs on the subject matter.[34]

---

[32] For a recent analysis of Fock's stance on FQM also in relation to Bohr's ideas see (Martinez 2019).

[33] The school of Geymonat, to which Tagliagambe belonged, had a manifest Marxist approach and was to have a lasting influence on the philosophy of science in Italy. However, the new generation of physicists who revived the FQM in the 1970s, harshly criticized their approach (see section 3.4). This likely prevented the work of Tagliagambe (1972) to become influential among the physicists sensitized about FQM.

[34] A comprehensive bibliography (encompassing 362 references) of the Italian studies on FQM up until 1985 can be found in (Benzi 1988), and in the rest of this section we refer the reader to that work for the missing references (see https://link.springer.com/chapter/10.1007/978-94-009-2947-0_20).

A number of other physicists, who were not active in the political struggles that inspired this revival, also performed research on FQM in the following years, often acquiring an international reputation. This is the case –just to mention a few notable ones– of Enrico Beltrametti and Gianni Cassinelli, in Genova, and Maria Luisa Dalla Chiara (a philosopher) and Toraldo di Francia, in Florence. They all contributed a great deal, eventually becoming international authorities, to an approach to FQM called "quantum logic". This subfield of FQM was pioneered by Garrett Birkhoff and John von Neumann (1936), and was experiencing a renewal in the 1960-70s. They attracted international attention, being praised by the eminent philosopher Bas van Fraassen. The latter even co-organized, with Beltrametti, a conference in Erice on "Current issues on quantum logic" (Beltrametti & van Fraassen 1981).

It ought to be remarked that the "school of Milano", mostly under the leadership of Prosperi, continued a prolific production of publications on FQM, after the aforementioned initial period in the 1960s, with new scholars such as Alberto Barchielli, Ludovico Lanz, and Pietro Bocchieri, some of whom are still active in this or closely related fields.

5. **Concluding remarks**

In this chapter, we have analyzed the contingent conditions that led to a revival of the research in FQM, in Italy at the turn of 1970s. We drew a connection between its origins and the social-political struggles for change of the '68 left-wing movements, which also involved a young generation of physicists. What was remarkable of the Italian case study is that prime academic institutions, such as the Italian Physical Society, represented a stepping stone for these young radical physicists to revolutionize the sensitivity towards an unconventional research field like FQM. These physicists criticized the structures of science (such as the military and industrial practices of the so-called *Big Science* in high energy physics) and regarded science as yet another manifestation of the capitalistic character of modern society. They thus saw in FQM a natural starting point to dismantling the certainties of contemporary physics and thus opening new room for a radical change in the practices and contents of physics and science. While most of the initial goals of the Italian researchers turned out to be unattainable – such as the sought breakdown of quantum mechanics by means of Bell's theorem– dozens of Italian physicists became sensitized towards the fundamental issues of quantum physics: They published altogether hundreds of papers on quantum foundations between the end of the 1960s and the 1980s and this arguably helped a great deal in creating the conditions that made FQM an established research field of physics.

**Acknowledgements**

I wish to express my gratitude to Christian Joas, Stefano Osnaghi, Martin Renner and Joshua Morris for their comments which greatly improved this chapter. I am moreover thankful to Luigi Romano for having shared with me some of his findings on archival material about Franco Selleri.

**References**

Andrade e Silva, J, Selleri, F & Vigier, JP 1983, 'Some Possible Experiments on Quantum Waves', *Lettere al Nuovo Cimento*, no. 36, pp. 503-508.

Aronova, E & Turchetti, S 2016, *Science Studies during the Cold War and Beyond*, Palgrave Macmillan, US.

Baracca, A, Bergia, S, Bigoni, R & Cecchini, A 1974, 'Statistics of observations for proper and improper mixtures in quantum mechanics', *Rivista del Nuovo Cimento*, no. 4, p. 169.

Baracca, A, Bergia S & Restignoli M 1974, 'On the comparison between quantum mechanics and local hidden variable theories: Bell's type inequality for multi-valued observables', in Mitra, N, Slaus, I, Bhasin, V & Gupta, V (eds.), *Proceedings of the International Conference on Few Body Problems in Nuclear and Particle Physics*, North-Holland, Amsterdam.

Baracca, A, Bergia, S & Del Santo F 2017, 'The origins of the research on the foundations of quantum mechanics (and other critical activities) in Italy during the 1970s', *Studies in History and Philosophy of Modern Physics* no. 57, pp. 66-79.

Baracca, A, Bohm, D, Hiley, B & Stuart, A 1975, 'On some notions concerning locality and nonlocality in the quantum theory', *Il Nuovo Cimento* no. 28B, p. 435.

Baracca, A, Bergia, S, Cannata, F, Ruffo, S, & Savoia, M 1977, 'Is a Bell-Type Inequality for Nondicotomic Observables a Good Test of Quantum Mechanics?', *International Journal of Theoretical Physics*, no. 16, p. 491.

Baracca, A & Del Santo F 2017, 'La giovane Generazione dei fisici e il rinnovamento delle scienze in Italia negli anni Settanta', *Altronovecento: Ambiente, Tecnica, Società*, no. 34.

Baracca, A, Cornia, A, Livi, R, & Ruffo, S 1978, 'Quantum mechanics, first kind states and local hidden variables: three experimentally distinguishable situations', *Il Nuovo Cimento B*, no. 43, p. 65.

Bedford, D & Selleri, F 1985, 'On Popper's new EPR-experiment', *Lettere al Nuovo Cimento*, vol. 42, no. 7, pp. 325-328.

Bell, JS 1964, 'On the Einstein Podolsky Rosen paradox', *Physics Physique Fizika*, vol. 1, no. 3, p. 195.

Bell, JS 1987, 'Are there quantum jumps?', in Kilmister CW (ed.), *Schrödinger, Centenary of a Polymath*, Cambridge University Press, Cambridge.

Bellone, E, Geymonat, L, Giorello, G & Tagliagambe, S 1974, *Attualità del Materialismo Dialettico*, Editori Riuniti, Roma.

Beltrametti, EG & Van Fraassen, BC 2012, *Current issues in quantum logic*, Springer Science & Business Media.

Benzi, M 1988, 'Italian Studies in the Foundations of Quantum Physics. A Bibliography (1965–1985)', In (Tarozzi & van der Merve 1988).

Bergia, S, Cannata, F, Cornia, A & Livi, R 1980, 'On the Actual Measurability of the Density Matrix of a Decaying System by Means of Measurements on the Decay Products', *Foundations of Physics* no. 10, p. 723.

Bergia, S, & Cannata, F 1982, 'Higher-Order Tensors and Tests of Quantum Mechanics', *Foundations of Physics*, no. 12, p. 843.

Bergia, S, Cannata F & Monzoni, V 1985, 'Explicit Examples of Theories Satisfying Bell's Inequality: Do They Miss Their Goal Prior to Contradicting Experiments?', *Foundations of Physics* no. 15, p. 145.

Bergia, S, Cannata, F & Pasini, A 1988, 'Space Time Fluctuations and Stochastic Mechanics: Problems and perspectives', in Kostro, L, Posiewnik, A, Pycacz J, Zukowski, M (eds.), *Problems in Quantum Physics*, World Scientific, Gdansk.

Bergia, S, Cannata, F & Pasini, A 1989, 'On the possibility of interpreting quantum mechanics in terms of stochastic metric fluctuations', *Physics Letters*, no. 137A, p. 21.

Besson, V 2018, *L'interprétation causale de la mécanique quantique: biographie d'un programme de recherche minoritaire (1951–1964)*, Education, University of Lyon (Doctoral dissertation).

Bertlmann, R, & Zeilinger, A (eds.) 2013, *Quantum (un)speakables: from bell to quantum information*. Springer, Vienna.

Birkhoff, G & von Neumann, J 1936, 'The logic of quantum mechanics', *Annals of mathematics*, vol. 37, no. 4, pp. 823-843.

Bonsignori, F & Selleri, F 1960, 'Pion cloud effects in pion production experiment' *Il Nuovo Cimento*, vol. 15, no. 3, pp. 465-478.

Bohm, D 1952, 'A suggested interpretation of the quantum theory in terms of "hidden variables" I', *Physical review*, vol. 85, no. 2, p.166.

Caldirola, P 1961, Quantistica, Meccanica, entry in the *Enciclopedia Italiana*, III Appendix.

Caldirola, P 1965, 'Teoria della misurazione e teoremi ergodici nella meccanica quantistica', *Giornale di Fisica*, no. 6, pp. 228–237.

Caldirola, P 1974, *Dalla microfisica alla macrofisica*, Mondadori, Milano.

Caldirola, P & Loinger, A 1957, 'L'interpretazione della teoria quantistica', *Il Pensiero*, Milano.

Capasso, V, Fortunato, D & Selleri, F 1973, 'Sensitive Observables of Quantum Mechanics', *International Journal of Theoretical Physics*, no. 7, pp. 319-326.

Ciccotti, G, Cini, M, De Maria, M, Jona-Lasinio, G, Donini, E, et al. 1976, *L'Ape e l'Architetto: Paradigmi Scientifici e Materialismo Storico*. Feltrinelli, Milan.

Cini, M 1969, 'Il Satellite della Luna', *Il Manifesto (Rivista)*, September.

Cini, M, De Maria, M, Mattioli, G & Nicolò, F 1979, 'Wave packet reduction in quantum mechanics: a model of a measuring apparatus', *Foundations of Physics*, vol. 9, no. 7, pp. 479-500.

Cini, M 1983, 'Quantum theory of measurement without wave packet collapse', *Il Nuovo Cimento B*, vol. 73, no. 1, pp. 27-56.

Clauser, JF 1992, 'Early history of Bell's theory and experiment', in Black TD (ed.), *Foundations of Quantum* Mechanics, World Scientific, Singapore.

Del Santo, F 2018, 'Genesis of Karl Popper's EPR-like experiment and its resonance amongst the physics community in the 1980', *Studies in History and Philosophy of Modern Physics*, no. 62, pp. 56-70.

Del Santo, F 2019 'Karl Popper's forgotten role in the quantum debate at the edge between philosophy and physics in 1950s and 1960s', *Studies in History and Philosophy of Modern Physics*, no. 67, pp. 78-88.

Del Santo, F 2020, 'An Unpublished Debate Brought to Light: Karl Popper's Enterprise against the Logic of Quantum Mechanics', *arXiv* preprint, 1910.06450.

D'Espagnat, B 1965, *Conceptions de la physique contemporaine: les interprétations de la mécanique quantique et de la mesure*. Editions Hermann, Paris.

D'Espagnat, B (ed.) 1971, *Proceedings of the International School of Physics "Enrico Fermi", Foundations of Quantum Mechanics 1970,* Academic Press.

Falciglia, F, Garuccio, A & Pappalardo, L 1982 'Rapisarda's experiment: on the four-coincidence equipment "FOCA-2", a test for nonlocality propagation', *Lettere al Nuovo Cimento*, no. 34, pp. 1-4.

Finkbeiner, A 2006, *The Jasons: The secret history of science's postwar elite*, Penguin, USA.

Fortunato, D & Selleri, F 1976, 'Sensitive Observables in Infinite-Dimensional Hilbert Spaces', *International Journal of Theoretical Physics*, no. 15, pp. 333-338.

Fortunato, D, Garuccio, A & Selleri, F 1977, 'Observable Consequences from Second-Type State Vectors of Quantum Mechanics', *International Journal of Theoretical Physics*, no. 16, pp. 1-6.

Freire Jr, O 2004, Popper, 'Probabilidade e mecânica quântica', *Episteme* no. 18, pp. 103-127.

Freire Jr, O 2011, 'On the connections between the dialectical materialism and the controversy on the quanta' *Jahrbuch Für Europäische Wissenschaftskultur*, no. 6, pp. 195–210.

Freire Jr, O 2014, *The Quantum Dissidents: Rebuilding the Foundations of Quantum Mechanics (1950-1990)*, Springer, Berlin.

Freire Jr, O 2019, *David Bohm: A Life Dedicated to Understanding the Quantum World*. Springer Nature, Switzerland.

Gagliasso, E, Della Rocca, M & Memoli, R 2015, *Per una scienza critica, Marcello Cini e il presente: filosofia, storia e politiche della ricerca*, Edizioni ETS: Pisa.

Garuccio, A, Popper, KR & Vigier, JP 1981, 'Possible direct physical detection of de Broglie waves', *Physics Letters A*, vol. 86, no. 8, pp. 397-400.

Garuccio, A & Rapisarda, V 1981, 'Bell's inequalities and the four-coincidence experiment' *Nuovo Cimento*, no. 65°, p. 289.

Ghirardi, GC, Omero, C, Rimini, A & Weber, T 1978, 'The Stochastic Interpretation of Quantum Mechanics: a Critical Review', *Rivista del Nuovo Cimento*, vol. 1, no. 3, pp. 1-34.

Ghirardi, GC, Rimini, A & Weber, T 1976, 'Implications of the Bohm-Aharonov Hypothesis' *Il Nuovo Cimento*, no. 31B, p. 177.

Ghirardi, GC, Rimini, A & Weber, T 1984, 'A Model for a Unified Quantum Description of Macroscopic and Microscopic Systems Quantum Probability and Applications', in Accardi, L et al. (eds.), *Quantum Probability and Applications II*, Springer, Berlin.

Ghirardi, GC, Rimini, A & Weber, T 1986, 'Unified dynamics for microscopic and macroscopic systems', *Physical Review D*, vol. 34, no. 2, p. 470.

Ghirardi, GC & Weber, T 1979, 'On Some Recent Suggestions of Superluminal Communication through the Collapse of the Wave Function', *Lettere Nuovo Cimento*, no. 26, p. 599.

Ghirardi, GC 1988, 'Some Critical Considerations on the Present Epistemological and Scientific Debate on Quantum Mechanics', in (Tarozzi & van der Merwe 1988).

Ghirardi, GC, Pearle, P & Rimini, A 1990, 'Markov processes in Hilbert space and continuous spontaneous localization of systems of identical particles', *Physical Review A*, no. 42, p. 78.

Ghirardi, GC 2017 'Some reflections inspired by my research activity in quantum mechanics', *Journal of Physics A: Mathematical and Theoretical*, vol. 40, no. 12, p. 2891.

Gisin, N 1989, 'Stochastic Quantum Dynamics and Relativity', *Helvetica Physica Acta*, vol. 62, no. 4, p. 363-371

Jammer, M 1974, *The Philosophy of Quantum Mechanics: The Interpretations of Quantum Mechanics in Historical Perspective*, Wiley, New York.

Kaiser, D 2011, *How the hippies saved physics: science, counterculture, and the quantum revival*, WW Norton & Company, New York.

Kim, YH & Shih, Y 1999, 'Experimental realization of Popper's experiment: Violation of the uncertainty principle?', *Foundations of Physics*, vol. 29, no. 12, pp. 1849-1861.

Trischler, C & Kojevnikov, A (eds.) 2011, *Weimar culture and quantum mechanics: Selected papers by Paul Forman and contemporary perspectives on the Forman thesis*. World Scientific.

Herbert, N 1982, 'FLASH-A superluminal communicator based upon a new kind of quantum measurement', *Foundations of Physics*, no. 12, pp. 1171–1179.

Livi, R 1977, 'New Tests of Quantum Mechanics for Multivalued Observables', *Lettere Nuovo Cimento*, no. 19, p. 272.

Martinez, Jean-Philippe. 'Beyond Ideology: Epistemological Foundations of Vladimir Fock's Approach to Quantum Theory', *Berichte zur Wissenschaftsgeschichte*, vol. 42, no. 4, pp. 400–423.

Moore, K 2013. *Disrupting science: Social movements, American scientists, and the politics of the military, 1945-1975*. Princeton University Press, Princeton (New Jersey).

Mückenheim, W, Lokai, P & Burghardt, B 1988, 'Empty waves do not induce stimulated emission in laser media', *Physics Letters A*, vol. 127, no. 8, pp. 387-390.

Nutricati, P 1998, *Oltre I Paradossi della Fisica Moderna: I Fisici Italiani per il Rinnovamento di Teoria Quantistica e Relatività*, Dedalo, Bari.

Ortigoso, J 2018, 'Twelve years before the quantum no-cloning theorem', *American Journal of Physics*, vol. 86, no. 3, pp. 201-205.

Romano, L 2020 (forthcoming), *Franco Selleri and his contribution to the debate on Particle Physics, Foundations of Quantum Mechanics and Foundations of Relativity Theory*. Università degli Studi di Bari (Doctoral dissertation).

Rosenfeld, L 1965, 'The measuring process in quantum mechanics', *Suppl. Progr. Theor. Phys.*, pp. 222-231.

Selleri, F 1969a, 'On the wave function in quantum mechanics', *Lettere al Nuovo Cimento*, vol. 1, no. 17, pp. 908-910.

Selleri, F 1969b (unpublished), 'Quantum Theory and Hidden Variables', lectures held in Frascati (June-July), LNF – 69/75 CNEM-Laboratori Nazionali di Frascati.

Selleri, F 1972, 'A Stronger Form of Bell's Inequality', *Lettere al Nuovo Cimento*, vol. 3, no. 14, pp. 581-582.

Selleri, F & Tarozzi, G 1978, 'Nonlocal Theories Satisfying Bell's Inequality', *Nuovo Cimento*, vol. 48B, no. 1, pp. 120-130.

Selleri, F 1982a, 'Can an Actual Existence be Granted to Quantum Waves?', *Annales de la Fondation Louis de Broglie*, no. 7, pp. 45-73.

Selleri, F 1982b, 'On the direct observability of quantum waves', *Foundations of Physics*, no. 12, pp.1087-1112.

Selleri, F 1983. *Die Debatte um die Quantentheorie*, Springer-Verlag.

Tagliagambe, S (ed.) 1972, *L'Interpretazione materialistica della meccanica quantistica. Fisica e filosofia in URSS*, Feltrinelli, Milano.

Tarozzi, G & van der Merwe, A (eds.) 1985, *Open Questions in Quantum Physics*, Reidel Publishing Co, Dordrecht.

Tarozzi, G & van der Merwe, A (eds.) 1988, *The Nature of Quantum Paradoxes: Italian Studies in the Foundations and Philosophy of Modern Physics*, Kluwer Academic Publishers, Dordrecht.

van der Merwe, A, Selleri, F & Tarozzi, G (eds.) 1988, *Microphysical Reality and Quantum Formalism (Vol. 1 and 2)*, Kluwer Academic Publishers, Dordrecht.

Vitale, B (ed.) 1976, *The War Physicists*, Napoli.

Weiner, C (ed.) 1977, *Proceedings of the International School of Physics "Enrico Fermi", History of Twentieth Century Physics 1972*, Academic Press.

Wootters, W., Zurek, W. 1982, 'A single quantum cannot be cloned', *Nature* no. 299, pp. 802–803.